\documentclass[sigconf]{acmart}
\renewcommand\footnotetextcopyrightpermission[1]{}

\usepackage{graphicx}
\usepackage{hyperref}
\usepackage[utf8]{inputenc}

\usepackage{amsmath,amssymb,amsfonts}
\usepackage{algorithmic}
\usepackage{graphicx}
\usepackage{textcomp}
\usepackage{xcolor}
\usepackage{array}
\usepackage{xspace}
\usepackage{multirow}
\usepackage{makecell}
\usepackage{listings}
\usepackage{enumitem}
\usepackage{float}
\usepackage{listings}
\usepackage{tcolorbox}
\usepackage{subcaption}
\newtcolorbox{answerbox}{
  colback=white,   
  colframe=black,  
  sharp corners,   
  boxrule=0.8pt,   
  left=6pt,        
  right=6pt,       
  top=4pt,         
  bottom=4pt,      
  before skip=5pt,
  after skip=5pt, 
  fontupper=\normalsize\flushleft 
}
\newcolumntype{C}[1]{>{\centering\arraybackslash}m{#1}}

\AtBeginDocument{%
  }

\setcopyright{acmlicensed}
\copyrightyear{2018}
\acmYear{2018}
\acmDOI{XXXXXXX.XXXXXXX}
\acmConference[Conference acronym 'XX]{Make sure to enter the correct
  conference title from your rights confirmation email}{June 03--05,
  2018}{Woodstock, NY}
\acmISBN{978-1-4503-XXXX-X/2018/06}
\newcommand{\codename}{\mbox{\textsc{PDLogger}}\xspace}

\begin{document}

\title{\codename: Automated Logging Framework for Practical Software Development}




\author{Shengchen Duan}
\affiliation{
  \institution{Singapore Management University}
  \city{Singapore}
  \country{Singapore}
}

\author{Yihua Xu}
\affiliation{%
  \institution{Illinois Institute of Technology}
  \city{Chicago}
  \state{Illinois}
  \country{USA}
}

\author{Shen Wang}
\affiliation{
  \institution{Singapore Management University}
  \city{Singapore}
  \country{Singapore}
}

\author{Sheng Zhang}
\affiliation{%
  \institution{Hefei University of Technology}
  \city{Hefei}
  \state{Anhui}
  \country{China}
}

\author{Yue Duan}
\affiliation{
  \institution{Singapore Management University}
  \city{Singapore}
  \country{Singapore}
}

\renewcommand{\shortauthors}{Trovato et al.}

\begin{abstract}
Logging is indispensable for maintaining the reliability and diagnosability of modern software, yet developers still struggle to decide where and how to log effectively. Existing automated logging techniques focus on isolated sub-tasks—predicting a single log position, level, or message—and therefore cannot produce complete, high-quality log statements that reflect real-world practice in which multiple logs often appear inside one method. They also neglect deeper semantic dependencies among methods and consider only a narrow set of candidate variables, leading to superficial or incomplete logs. In this paper, we present \codename, the first end-to-end log generation technique expressly designed for practical, multi-log scenarios. \codename operates in three phases. (1) Log position prediction: block-type–aware structured prompts guide a large language model (LLM) to suggest candidate positions across all control-flow blocks of a method. (2) Log generation: backward program slicing supplies precise inter-procedural control‐ and data-dependency context, while an expanded variable extractor captures both member and external function expressions; the enriched prompt enables the LLM to emit a full log statement (position, level, message, variables). (3) Log refinement: level correction and context-sensitive deduplication prune false positives and redundant logs. We evaluate \codename on 3,113 log statements drawn from two widely used Java projects. Compared with the strongest prior systems, \codename improves log-position precision by 139.0\%, F1 by 69.2\%, level accuracy by 82.3\%, variable precision by 131.8\%, and message quality (BERTScore) by 65.7\%. The framework consistently performs well with different mainstream LLMs, demonstrating robustness and generality. \codename’s implementation is available as open source to foster future research and adoption.
\end{abstract}

\maketitle
\section{Introduction}
With the growing complexity and scale of software systems, logging has evolved from optional to a widely recognized and indispensable mechanism for ensuring software reliability and integrity. However, unreasonable logging practices not only compromise the readability of logs but also influence system performance\cite{b26}. To address this issue, numerous automated logging schemes have been proposed to help developers insert log statements more effectively. A typical logging statement consists of four key components: position, level, message, and variable. Most prior techniques target only individual components (e.g., predicting log positions \cite{b23, b61, b62, b63, b26}, levels \cite{b6,b7,b9}, or messages \cite{b10,b59,b60}) and are not able to generate complete logs, making them unsuitable for real-world development. 

Recent advances in large language models (LLMs) have attracted much attention in the field of automated logging. LLMs learn common logging patterns from massive code corpora, understand both natural language and code, and can infer a developer’s intent from surrounding context. This makes them potentially capable of generating concise, informative messages that follow project-specific conventions, incorporate relevant variables, and assign appropriate severity levels. As a result, LANCE~\cite{b17}, the first end-to-end log generation approach, is built on an LLM named T5~\cite{b11} and is capable of inserting complete log statements into given code snippets. However, it relies solely on intra-method information, overlooking contextual information from both callers and callees. To address it, SCLogger~\cite{b28} enhances log quality through static scope expansion, style adaptation, and context-aware prompt construction. Nevertheless, SCLogger fails to incorporate semantic dependency information, often generating superficial error descriptions that lack insight into the root cause.  UniLog \cite{b5} leverages LLM with line‑position‑sensitive in‑context learning prompts to predict a complete log without fine‑tuning; however, it likewise fails to capture deeper semantic dependencies. 

\vspace{4pt} \noindent \textbf{Limitations.} To sum up, we identify three key limitations in existing state-of-the-art approaches, which will be further illustrated with real-world examples in Section~\ref{sect:motivation}.

First, none of the existing approaches, by design, handle multiple log generation scenarios within a method. In modern software systems, it is extremely common for a single method to contain multiple logging statements, which are essential for enhancing system observability and facilitating fault diagnosis. Our analysis of the LogBench-O benchmark~\cite{b15} reveals 6,849 log statements across 3,870 methods, resulting in an average of 1.77 log statements per method. However, state-of-the-art automated logging techniques \cite{b31,b5,b28} are designed and assessed under the assumption that each method requires at most one log statement. This limitation severely hinders their applicability in real-world development scenarios. Therefore, it is imperative to develop techniques that support multi-log generation within a single method.

Second, state-of-the-art logging techniques lack sufficient semantic dependency context. Among the existing methods, only SCLogger~\cite{b28} attempts to incorporate contextual information; however, its context is limited to a small subset of randomly selected callers and callees (within two hops). As a result, the retrieved context is often incomplete and semantically irrelevant or misleading, rendering low-quality log generation. 

Third, all existing techniques suffer from a limited scope of logging variables. Recent studies, such as SCLogger, incorporate the member functions and attributes of the class that contains the given method into its log variable list. However, it overlooks the case where non-member functions or function expressions are used directly as logging variables. Our empirical study on 100 randomly sampled logs from Apache Hadoop~\cite{b56} reveals that 13\% of the logs utilize non-member functions or function expressions as variables. This observation highlights that the limited scope of log variable lists will likely result in the omission of critical variables, thereby reducing the expressiveness and diagnostic value of the logs.

\vspace{4pt} \noindent \textbf{Our Approach.} To tackle these issues, we propose \codename, the first automated log generation technique that is designed to be practical in real-world software development. It generates log statements through three major phases, namely log position prediction, log generation, and log refinement. In the log position prediction phase, we first extract the boundaries of code blocks within a given method and annotate their start and end positions to facilitate better recognition by LLM. We design block-type-based structured prompts to query the LLM. This helps address the inability to perform multi-log prediction and mitigates the excessive logging problem that arises in such scenarios. Then \codename leverages the LLM to predict log positions for different block types independently. The predicted positions may still contain false positives, which will be addressed in the refinement phase. In the log generation phase, based on the positions predicted in the first stage, our system applies a static backward slicing technique~\cite{sasirekha2011program} to generate backward slices, which can capture the precise semantic dependencies of the code around the logging positions and address the second limitation. 

Then, in addition to existing logging techniques, we further perform analysis to identify a set of external functions that can be used to generate log variables, thereby addressing the issue of the limited scope of logging variables. These dependencies and function information are then integrated into updated prompts, which are passed to LLM to generate complete log statements. After that, \codename adopts two major refinement strategies, level refinement and deduplication, to enhance the accuracy of the generated logs' levels and to identify similar logs within specific log contexts, removing redundant ones. This reduces the number of false-positive logs in the program, thereby mitigating the risk of developers being misled and improving the utility of log information.

To evaluate \codename, we acquire a comprehensive dataset comprising 3,113 log statements from two popular open-source Java projects and conduct a thorough evaluation. The results show that \codename can vastly outperform state-of-the-art log generation techniques in both single-log and multi-log generation scenarios. Specifically, \codename outperforms the baseline approach by 139.00\% in log position precision, 69.20\% in F1 score, 82.30\% in logging level accuracy, 131.80\% in variable precision, and 65.70\% in BERTScore \cite{b54} for log message generation. Moreover, \codename consistently achieves strong performance across a variety of mainstream LLMs, demonstrating its generalizability.

The contributions of this paper are summarized as follows:
\begin{itemize}
    \item We present \codename, an automated log statement generation technique for practical use. By embedding semantic dependency context, \codename overcomes prior work that captures only surface‑level signals.

    \item We design a novel prompt structure to improve prediction accuracy in multiple log generation scenarios. This structure can be integrated into other generative log generation methods and generalized to future LLM improvements.

    \item We investigate existing logs in real-world projects and summarize a deletion priority rule to reduce false positives. This rule is generalizable and can be applied to false-positive deduplication tasks in various log generation methods.

    \item We conduct a comprehensive evaluation of \codename on a dataset generated from public projects. The results show that \codename outperforms existing approaches and is adaptable to different LLMs.
    
    \item The source code of \codename is made publicly available\footnote{\url{https://github.com/qlbdsc/PDLogger}} to facilitate future research.
\end{itemize}

\section{Motivation}~\label{sect:motivation}

In this section, we highlight the motivation of our research by listing three major limitations of the existing logging techniques. 

\subsection{Lack of Multi-log Generation Capability}

Empirical evidence indicates that developers typically embed multiple log statements within a single method. However, state-of-the-art automated logging techniques \cite{b31,b5,b28} are designed and assessed under the assumption that each method requires at most one log statement. Although these techniques can, in principle, be extended to predicate multiple logs, such naive adaptations will likely introduce an excessive number of false positives. 

\vspace{-0.1in}
\begin{figure}[ht]
    \centering
    \includegraphics[width=\linewidth]{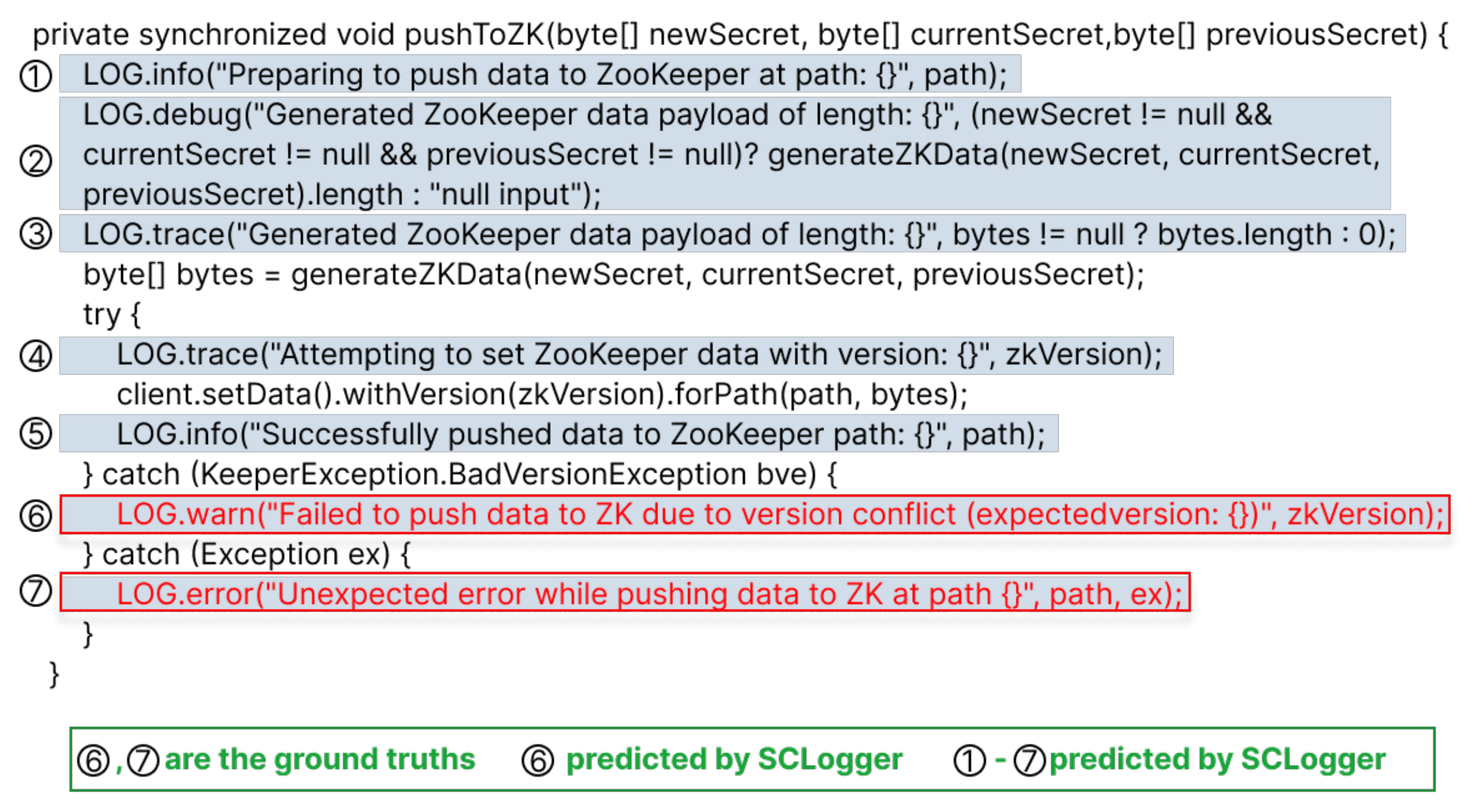}
    \vspace{-0.15in}
    \caption{Multi-Log Generation Issues}
    \vspace{-0.1in}
    \label{fig0}
\end{figure}

We illustrate this limitation with an example method from the ActiveMQ~\cite{b41} project, shown in Figure~\ref{fig0}. The function contains two logging statements written by the developer, which we treat as the ground truth and mark in red (6 and 7). The vanilla state-of-the-art technique, SCLogger, can only generate one log for the method (6), and thus fails to support multi-log generation. To accommodate multiple log cases, we adapt SCLogger~\cite{b28} by revising its LLM prompt to elicit predictions of several log statements, resulting in a new variant that we designate as \textit{Multi-SCLogger}. It generates seven logs — two true positives and five false positives (1-7), yielding 100\% recall but only 28.6\% precision. The results show that simply modifying existing techniques to support multi-log generation is ineffective in practice, as it leads to performance degradation and significant data redundancy. Consequently, we need a system that is specifically designed to handle multiple-log scenarios. 



\subsection{Insufficient Semantic Dependency Context}

In modern software systems, a method’s execution often depends on interactions with other methods~\cite{b37}. Although approaches like SCLogger~\cite{b28} extend context from the method level by including randomly selected caller and callee methods within two hops, such random selection will likely fail to capture complete relevant information for log generation, resulting in inaccurate logs. To improve, it is crucial to analyze the precise contextual information by incorporating control and data dependencies, which reveal execution prerequisites and upstream influences vital for understanding the semantics and execution conditions of the target line~\cite{b38}.

\begin{figure}[htbp]
    \centerline{\includegraphics[width=\linewidth]{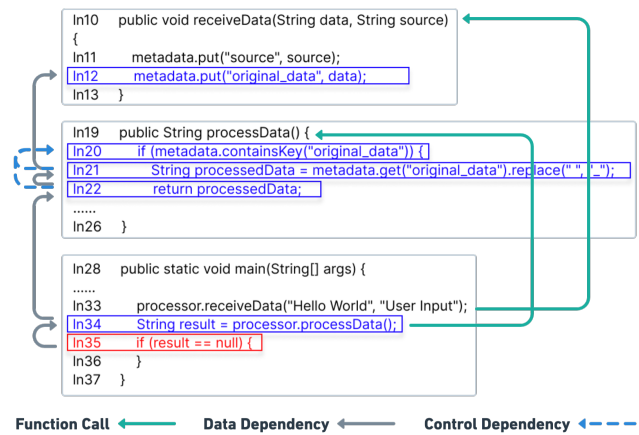}}
    \caption{Semantic Dependency Tracking. The red-highlighted line marks the starting statement of the backward slice. 
    Blue-highlighted lines denote statements reached through data and control dependency tracing.}
    \label{fig1}
\end{figure}

Figure~\ref{fig1} highlights the importance of extracting statement-level semantic dependencies. When inserting a log after \texttt{if (result == null) (ln.35)}, SCLogger's random sampling may only capture the \texttt{receive\_data} function (ln.10), offering little insight to infer the actual cause of the condition. In contrast, tracing control and data dependencies reveals that the condition at ln.35 stems from ln.34, which depends on \texttt{processor.process\_data()} at ln.22, itself relying on earlier lines. Therefore, the root cause of the condition \texttt{result == null} is the absence of the value associated with the \verb|original_data| key in \verb|metadata|.

Without the dependency information of the target line, it is nearly impossible to infer that the condition holds due to the missing \texttt{original\_data}. Moreover, if \texttt{DataProcessor()} is randomly selected during context expansion, the extracted information may be irrelevant to the log, reducing log utility. More critically, if a standard library function is chosen, its function body is typically inaccessible, offering no meaningful context. Therefore, to generate high-quality and contextually relevant log statements, it is necessary to obtain more precise contextual information from the statement level by leveraging semantic dependencies.

\subsection{Limited Log Variable Scope}

Log statements often contain variables to enhance log readability, improve debugging effectiveness, and increase the utility of automated analysis~\cite{b18,b40}. For instance, \texttt{"LOG.info("Starting with address: {}",getDatanodeAddress());"} includes a function call to retrieve and output the actual runtime address of the current node. The return value of this function enables developers to quickly identify key information, such as which specific node initiates the startup process. Prior studies~\cite{b12,b13,b14} have shown that, compared to purely textual logs, developers prefer log statements that contain program state variables, as they facilitate issue localization and comprehension of system behavior. However, when considering potential logging variables, existing techniques (e.g., SCLogger~\cite{b28}) focus only on member functions and local variables, but overlook cases where non-member functions and function expressions should also be considered.

\vspace{-0.15in}
\begin{figure}[h]
    \centering    
    \includegraphics[width=\linewidth]{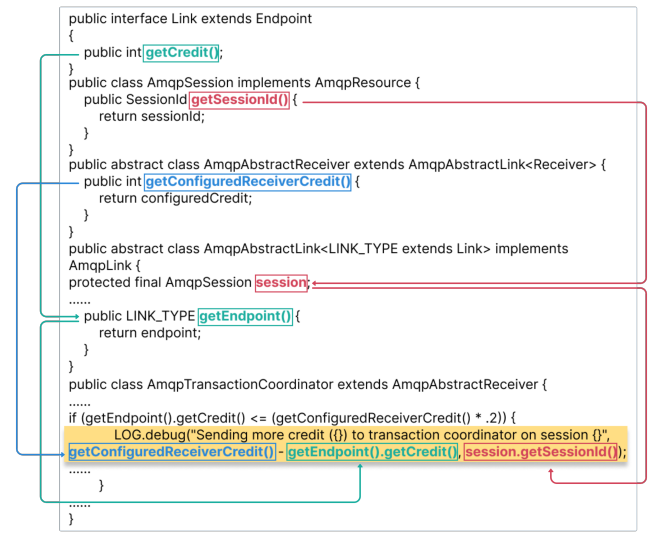}
    \caption{Log Variable Origins}
    \vspace{-0.1in}
    \label{fig2}
\end{figure}

Figure~\ref{fig2} illustrates an example from the \textit{ActiveMQ}~\cite{b41}project, where the log statement includes both a function call and arithmetic operations involving multiple functions as its variables. The log is intended to record the status of sending additional credit to a transaction coordinator within a particular session. To achieve this, the log message integrates information derived from non-member functions defined outside the current class. These pieces of information are essential for developers to understand the system’s behavior regarding resource allocation and transaction coordination during message processing. Although such omissions may not significantly impact traditional variable-level metrics in log evaluation, they can substantially degrade the accuracy of log-based program analysis and anomaly detection. This limitation can be addressed by extending non-member functions extracted from semantic dependency contextual information into the variable list during log generation.
\section{System Design and Implementation}

\subsection{Overview}

To address the aforementioned limitations, we propose \codename, an automated LLM-based log generation framework that can be practically applicable to real-world software development.

\begin{figure}[htbp]
    \centering
    \includegraphics[width=\linewidth]{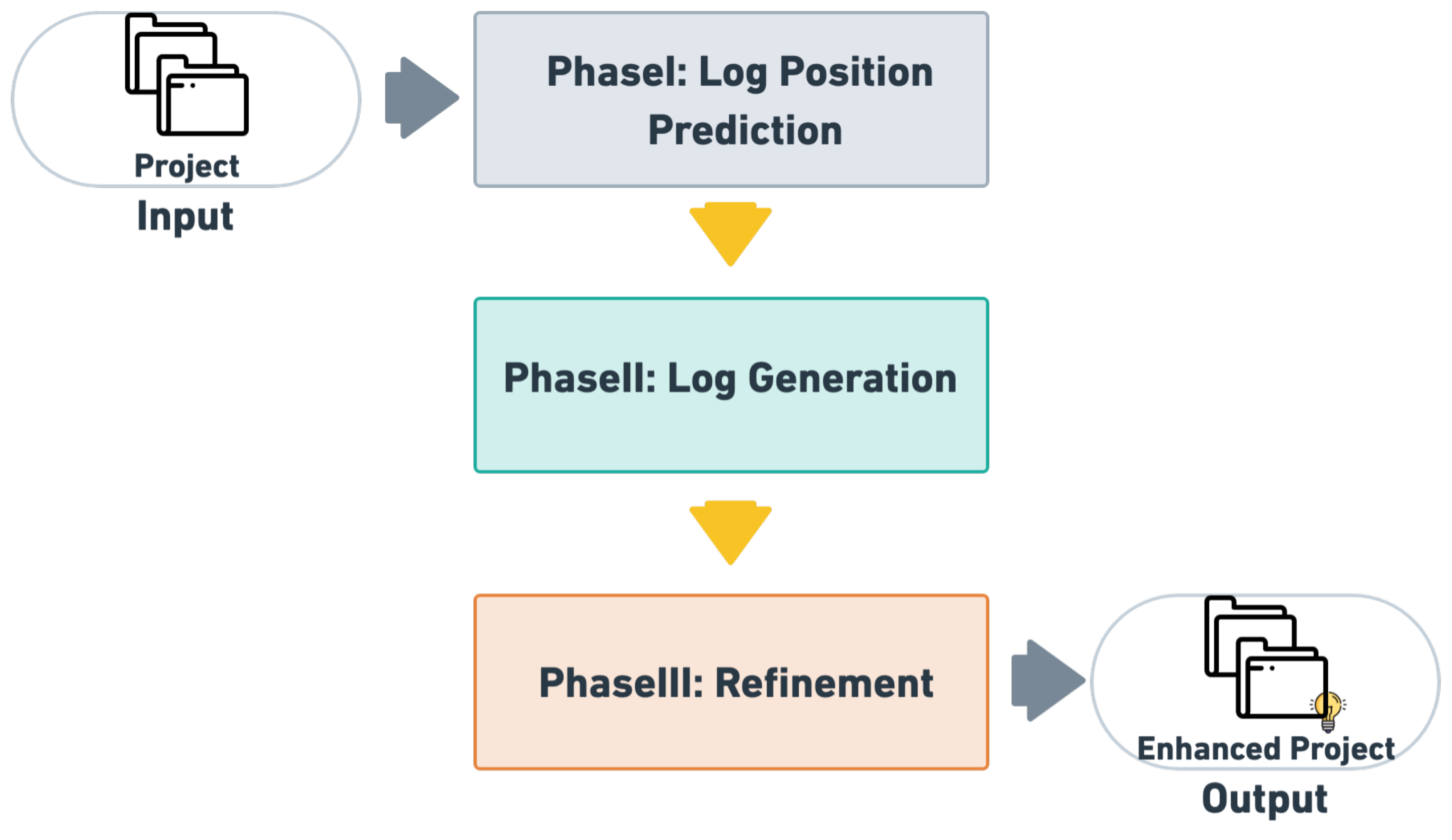}
    \caption{Overview of \codename}
    \label{fig3}
\end{figure}

Figure~\ref{fig3} illustrates the overview of our system, which takes a project codebase as input and outputs an enhanced project with log statements generated. The entire system consists of three major phases: log position prediction, log generation, and refinement. Given a project codebase, we generate logs for each method by following these three phases. Specifically, in the log position prediction phase, we identify the start and end positions of different code blocks, following a prior work~\cite{b23} that defines different code block types within a function and studies where log statements are most likely to exist. Then, we annotate the method with the block information to form a block-type-based structured prompt, which is fed into the LLM to predict log positions.


Then, in the log generation phase, we use the predicted log positions to perform backward slicing, incorporating both data and control dependencies information into the prompt. We then extract the variable list and enrich it with detailed function-level information. This enriched semantic context is combined with the context extracted in the previous phase, along with CoT~\cite{b43} reasoning, to form a new prompt. This prompt is fed into the LLM to obtain the final logging statements. 

Since the log level reflects the importance of a logging statement~\cite{b44,b45}, in the refinement phase, we conduct a level refinement to improve the accuracy of log level prediction. Finally, to reduce false positives, \codename introduces a unique deduplication process, producing a filtered set of logging statements. The final output is the final version of a project, containing the generated log statements with redundant entries removed.

\subsection{Log Position Prediction}
To extract a precise semantic dependency context, we need to find an accurate starting point. Therefore, the first step in \codename is to predict log positions. To do so, for the given project, we perform static analysis for each method and identify all code blocks. Then, we annotate the method with the start and end positions of the code blocks to create an annotated method. This step generates customized prompts for the LLM to ensure high precision and low false positive rates when predicting log positions, especially in the multi-log generation scenario. This step is critical because if the position prediction is wrong, the subsequent predictions of message, level, and variables will all become meaningless.

\begin{figure}[htbp]
    \centering
    \includegraphics[width=\linewidth]{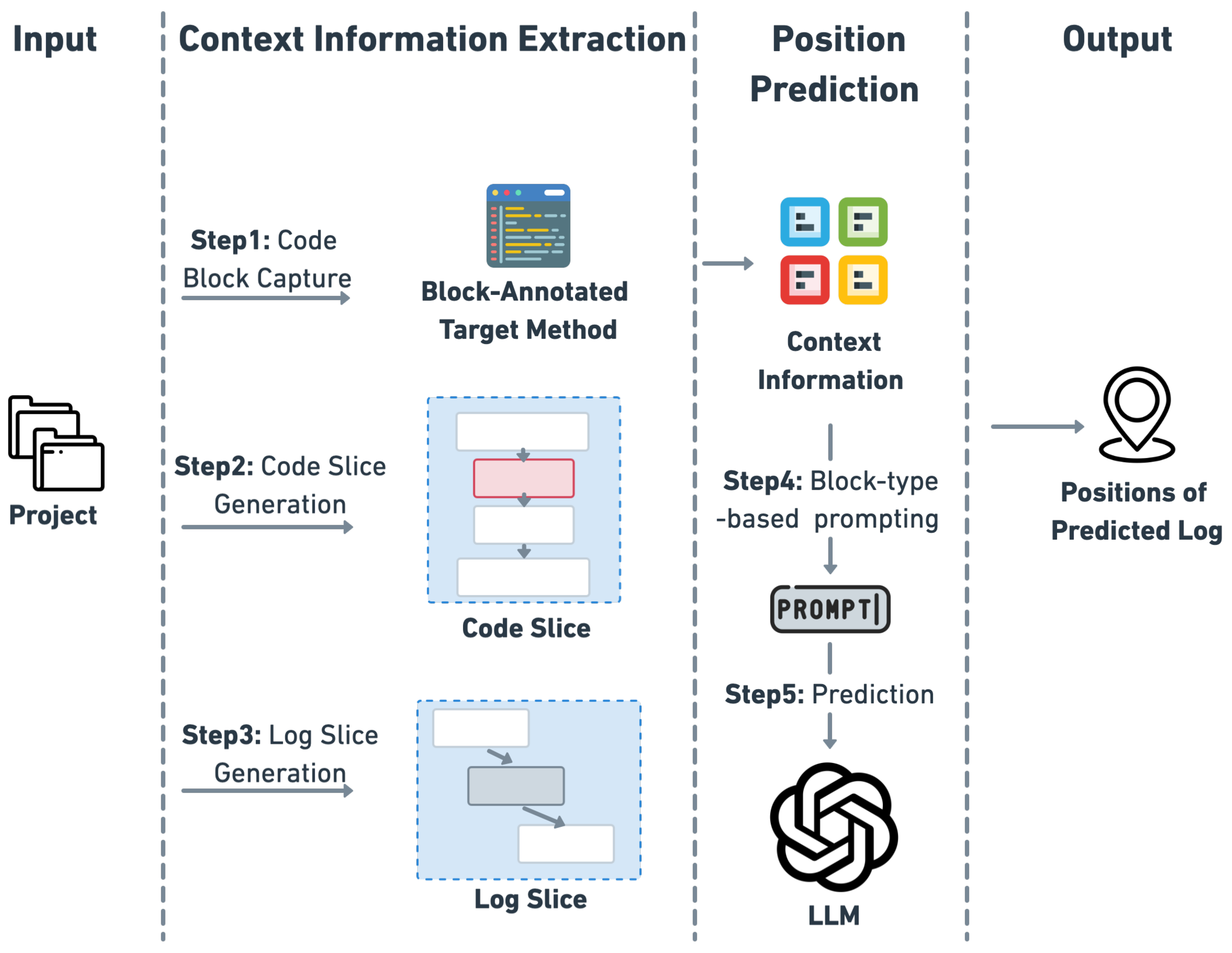}
    \caption{Log Position Prediction Workflow}
    \label{fig4}
\end{figure}

\vspace{4pt} \noindent \textbf{Block-type-based Structured Prompt Construction.} As shown in Figure~\ref{fig4}, \codename categorizes the blocks within a method into four types: branch blocks, try-catch blocks, loop blocks, and method definition blocks\cite{b23}. For each type, \codename counts the number of corresponding blocks, assigns unique identifiers, and marks the start and end lines of each block. Then, a corresponding prompt is constructed for each block type and fed into the LLM, with the number of queries corresponding to the number of such blocks with the target method.

\begin{table}[]
\caption{Block-Type-Based Heuristic Rules.}
\begin{tabular}{|>{\centering\arraybackslash}m{0.25\linewidth}|p{0.6\linewidth}|}

\hline
\textbf{Block Type}     & \multicolumn{1}{c|}{\textbf{Rule}}                                                                                           \\ \hline
Branch Block            & Add logs after events that directly affect program behavior and outcomes, such as exception handling.                        \\ \hline
Try-catch Block         & For cases with multiple catch blocks, evaluate each catch block individually to determine whether a log should be generated. \\ \hline
Looping Block           & Avoid printing logs that remain unchanged across multiple iterations of the loop.                                            \\ \hline
Method Definition Block & Insert logs into methods that are responsible for recording time-related information.                                        \\ \hline
\end{tabular}
\label{rules}
\vspace{-0.1in}
\end{table}

To enhance the precision of position in multi-log prediction while reducing the false positive rate of position, each prompt for a different type of block incorporates a set of heuristic rules, shown in Table~\ref{rules}. These rules are derived through our in-depth review of prior studies~\cite{b13,b26,b50,b51,b52} and our empirical analysis of log placement principles, particularly the structural and semantic cues that govern when and where logs should be generated. Importantly, these block-specific prompting strategies can be readily generalized to other generative log synthesis frameworks. The four prompt templates have been made publicly available on our GitHub repository.

\vspace{4pt} \noindent 
\textbf{Position Prediction.} To extract backward slices with accurate semantic dependencies, this step is designed to predict log positions. We first extract inter-procedural information comprising a code slice and a corresponding log slice. The code slice originates from a method and includes randomly selected callers or callees within two hops. The log slice contains log statements found in the code slice, arranged in the order of execution. As shown in Figure~\ref{fig4}, the extracted inter-procedural information is then combined with the block-type-based prompting to form a unified prompt, which is fed into the LLM to obtain log position predictions.

\subsection{Log Generation}
\begin{figure}[htbp]
    \centering
    \includegraphics[width=\linewidth]{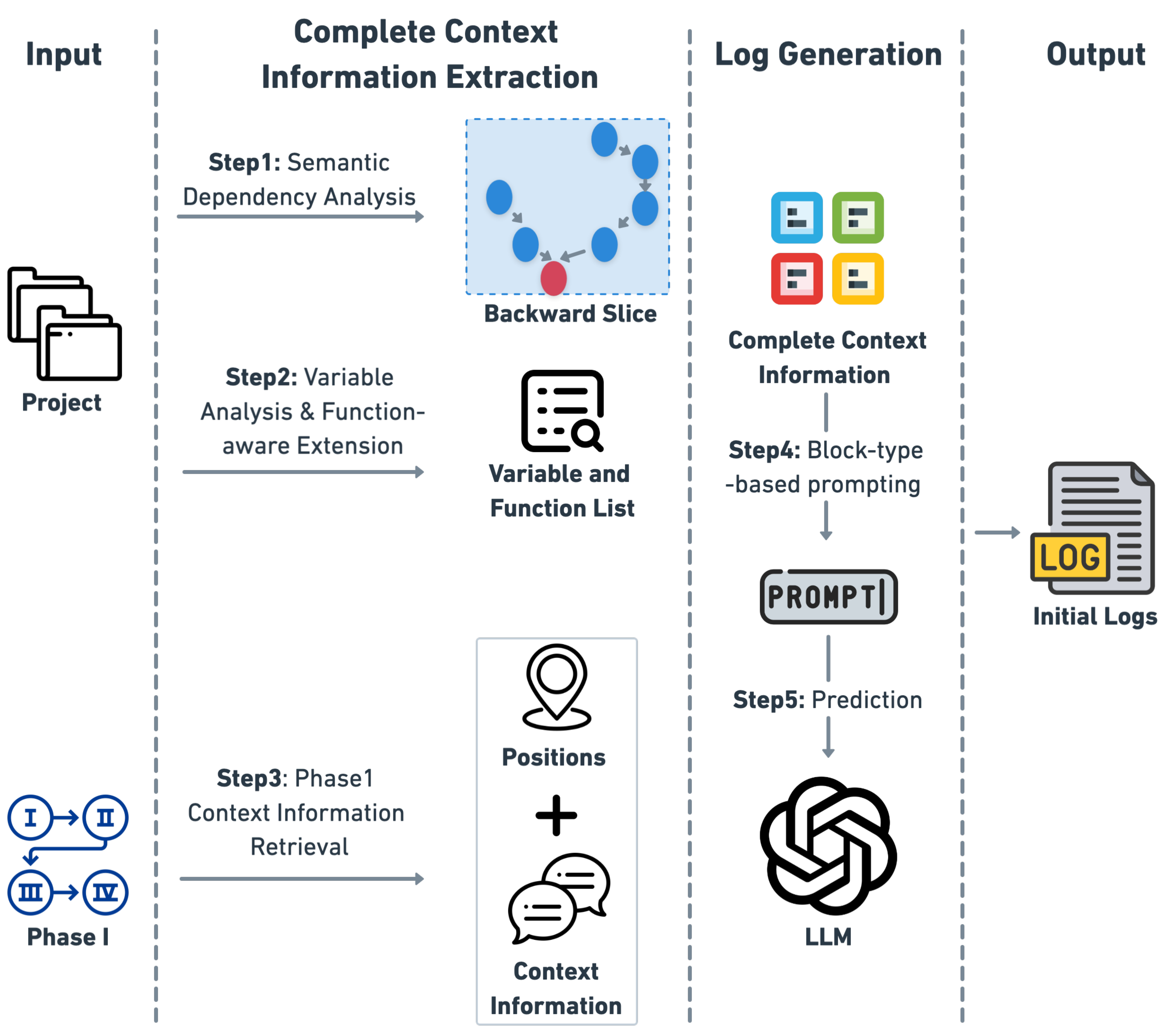}
    \caption{Log Generation Workflow}
    \label{fig5}
\end{figure}

After predicting the log positions, the next step for \codename is to generate log statements. As shown in Figure~\ref{fig5}, based on the predicted log positions within the target method, we first extract the statements corresponding to the line numbers and perform semantic dependency analysis to extract the contextual information needed for log statement generation. Further, we perform variable analysis and perform function-aware extension to enrich the potential variable list for our logs. Eventually, combined with the block-type-based prompting from Phase I, we query the LLM to produce an initial set of logs.

\vspace{4pt} \noindent \textbf{Semantic Dependency Analysis.} We define the precise semantic dependency information as the combination of control dependencies and data dependencies, which are fundamental in understanding how code statements in a program relate to each other. To extract such information, we perform backward slicing~\cite{sasirekha2011program} to generate backward slices based on the log positions predicted. To accommodate the input length limitations of LLMs, we constrain the backward slice to a maximum of seven hops. We define each "hop" as a traversal across a semantic dependency or control dependency edge. However, a key challenge arises in practice: for specific statements, the generated backward slice may be empty or meaningless. To address this, \codename iteratively selects preceding lines until reaching beyond the start of the current block. Specifically, for branch blocks, if the predicted position falls within an else block and no suitable line is found before exceeding the start of the else block, we instead select the corresponding if condition statement as the starting point of the backward slice. This is based on the observation that the if condition and the else block typically express opposite branches of the same logical judgment, often sharing the same set of conditional parameters~\cite{b47}. 


\vspace{4pt} \noindent \textbf{Variable Analysis and Function-aware Extension.} As highlighted in the motivation study, merely including member functions and class variables is insufficient to fully meet the practical requirements, as certain log statements rely on the return values of external functions as logging variables. Therefore, building upon SCLogger's approach\cite{b28} to expanding the variable scope, we further summarize the information of functions that are likely to appear as logging variables.

We first adopt the variable sets \( v \) defined in SCLogger, where \( v \in V_p \cup V_m \cup V_c \cup V_s \cup V_i \)\cite{b48}. Here, \( V_p \) denotes the set of parameters, \( V_m \) denotes the set of local variables, \( V_c \) denotes the set of class member variables, \( V_s \) denotes the set of static variables, and \( V_i \) denotes the set of inherited variables.
Within the context of the target method, we define functions sets  \( f \in F_m \cup V_i \cup V_d \cup V_l \cup V_s \) as follows:
\begin{itemize} [leftmargin=*]
    \item \textbf{\( F_m \)}: The set of member methods defined in the current class.
    \item \textbf{\( F_i \)}: The set of methods inherited from the parent class.
    \item \textbf{\( F_d \)}: The set of default methods from implemented interfaces.
    \item \textbf{\( F_l \)}: The set of lambda expressions or function variables declared within the current method.
    \item \textbf{\( F_s \)}: The set of statically imported methods from the parent or current class.
\end{itemize}

During the prediction of logging statements, the model should attend not only to the variable sets, but also to function variables \( f \) that belong to one or more of the above-defined function sets. Consequently, the context provided for logging variable selection should encompass both variables and function-derived values, collectively denoted as \( f \cup v \), which are treated as candidate logging variables and incorporated into the available variable list.

\subsection{Log Refinement}

After the previous step, \codename has already generated log messages that include all components. However, to better handle multi-log scenarios in practice, we need to perform log refinement to improve the quality of the logs as well as to reduce redundancy. 

\begin{figure}[htbp]
    \centering
    \includegraphics[width=\linewidth]{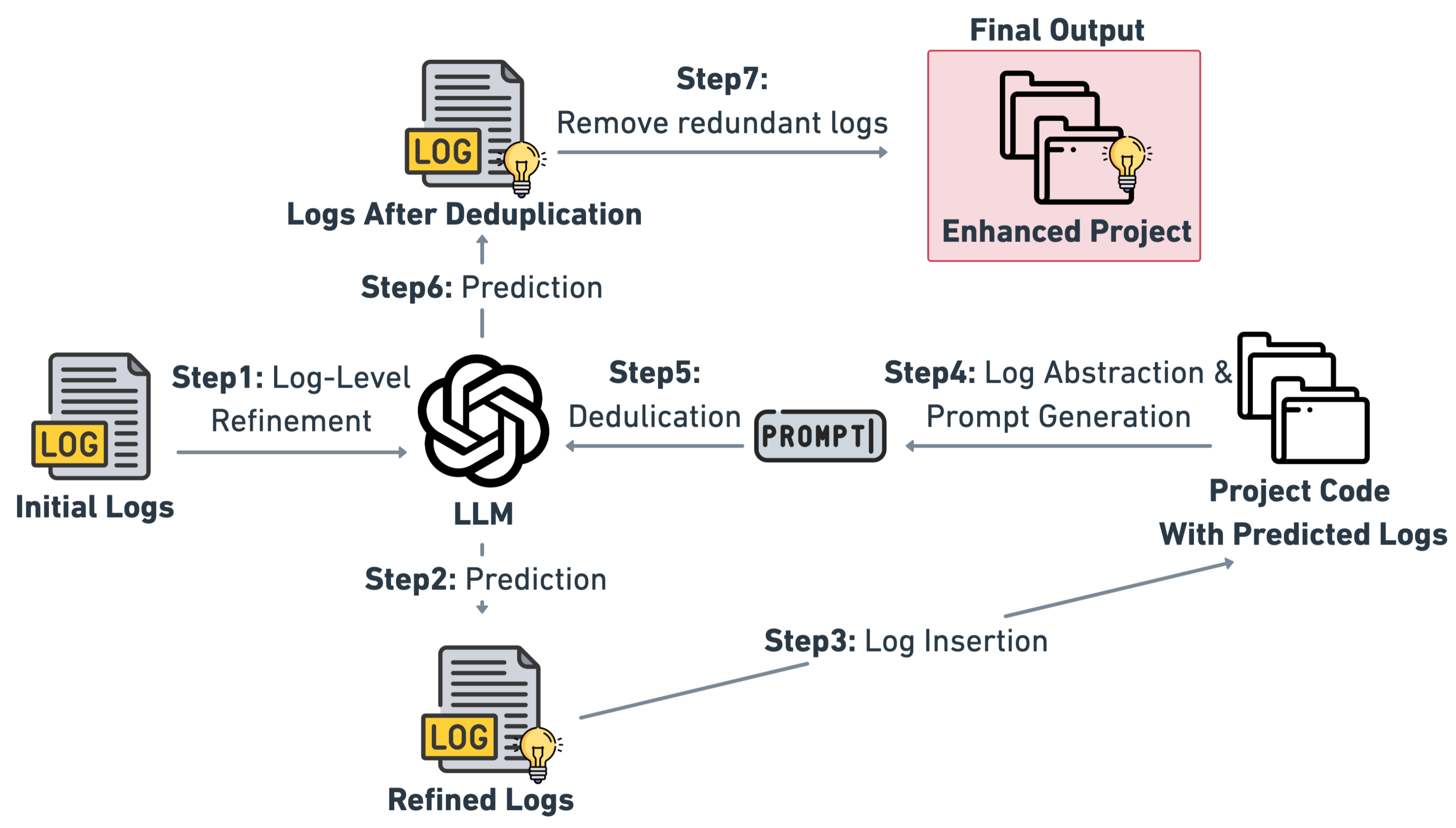}
    \caption{Log Refinement Workflow}
    \label{fig10}
\end{figure}

\vspace{4pt} \noindent 
\textbf{Log Level Refinement.}
As shown in~\ref{fig10}, we refine the initially predicted log levels before inserting the predicted log statements into the project. Log levels are essential for distinguishing the verbosity and severity of log messages, enabling developers to control log output across different environments, facilitate efficient debugging, and support runtime monitoring and alerting. Based on our investigation and some prior studies~\cite{b7,b49}, we identify five key factors that contribute to log level determination: (1) the content of the log message, (2) the method in which the log resides, (3) the semantic explanation of the log message, (4) the function of the block containing the target log, and (5) the total number of lines in that block. According to this information, we extract these five types of information along with the originally predicted log level and feed them to the LLM, which then decides whether the original log level requires adjustment. If so, it outputs the adjusted log level; otherwise, we retain the original one. Experimental results across different backbone models demonstrate that our proposed log-level refinement approach exhibits strong generalization ability. The full prompt used in this process has been made publicly available in our GitHub repository.

\vspace{4pt} \noindent 
\textbf{Deduplication.} The deduplication step is designed to reduce the false-positive rate of the predicted log statements, thereby mitigating the performance degradation that can arise from excessive logging. To this end, we first insert the generated logs into the given project codebase, and then construct a \textit{Code Context Bundle} defined as the target method together with the caller and callee methods within one hop, and eliminate redundant logs by examining whether the bundle already contains outputs that are semantically equivalent to the target log.

To avoid the line number shifts that would occur if logs were inserted in an arbitrary order, we first group the predicted logs by their corresponding file. For each code file, we insert all logs in descending order of line number, from the largest to the smallest. Note that we insert each generated log statement at the line immediately following the predicted position. To handle cases where some developers may split a single statement across multiple lines for readability or formatting purposes, we analyze the abstract syntax tree (AST) of the code to accurately identify the termination point of the complete conditional expression, thereby enabling precise insertion of log statements. Finally, we get a project augmented with the generated logs in the correct positions.

Then we perform the deduplication. To maximize coverage of potential semantic redundancies, we identify five situations in which a predicted log can be considered redundant; each is addressed below. (1) Overlap with throw messages. A log is removed if it conveys the same message as a throw statement, provided the exception is caught and its message printed, or it propagates uncaught to main(). For string-based exceptions, inter-procedural data-flow analysis enables the tracing of variable origins to compare messages more precisely.
(2) Contradictory Logs in an if-else Block. For logs in opposite branches of a \verb|if-else| block, we retain the one with a higher level. If levels match, we keep the else-branch log, as execution of the \verb|else| branch often signals an anomalous path of control.
(3)Start–End Log Pairs. We identify pairs of logs that are logically sequential, where the first log denotes the start of a process (\verb|start_log|) and the second denotes its completion (\verb|end_log|). If a (\verb|start_log|) is post-dominated by its (\verb|end_log|), the former is removed. This rule excludes trace-level logs, which capture fine-grained execution.
(4) Duplicate Semantics with Shared Variables. For semantically identical logs sharing a variable \verb|var_common|, static analysis checks for reassignment between them. If reassigned, the earlier log is removed; otherwise, the later log is removed. Without shared variables, we retain the log nearer to meaningful code (e.g., method calls, exceptions).
Applying these five language-agnostic rules yields the final Augmented Project, which can be adapted to any project because it relies solely on static analysis and semantic comparison.
\section{Evaluation}
In this section, we conduct a comprehensive evaluation on the effectiveness of \codename. Specifically, we aim to answer four essential research questions. 

\begin{itemize}[leftmargin=*]
    \item \textbf{RQ1:} How effective is \codename in the single-log generation scenario?
    
    \item \textbf{RQ2:} How effective is \codename in the multi-log generation scenario?
    
    \item \textbf{RQ3:} What is the impact of different components of \codename?
    
    \item \textbf{RQ4:} How generalizable is \codename on different backbone LLMs?
\end{itemize}

\subsection{Experiment Setup}
\vspace{4pt} \noindent 
\textbf{Dataset Selection.}
To evaluate \codename, we randomly select  3,113 log statements from two popular projects, Apache Hadoop 3.4.1~\cite{b56} and Apache ActiveMQ 5.18.7~\cite{b41}. These logs are distributed across 914 methods, with up to 13 log statements per method, representing real-world software development scenarios. For each sampled method, we extract the identifiers of its package, class, and method names, and locate all log statements that invoke two mainstream logging utilities, Log4j~\cite{b57} and SLF4J~\cite{b58}. We record each log statement, along with its corresponding line number in the source code, and use this information as the ground truth.

\vspace{4pt} \noindent 
\textbf{Baseline Techniques.}
In a single-log prediction setting, where exactly one log is randomly removed from each target method, the task is to recover this single log. We adopt SCLogger~\cite{b28}, the first contextualized logging-statement generation approach that exploits inter-method static contexts, as our primary baseline. We also include the first deep-learning-based complete log generation work, LANCE~\cite{b17}, and its successor, LANCE 2.0~\cite{b31}, as baselines. Approaches that target only specific sub-components of log generation are not considered.

In the multi-log prediction setting, we adapt SCLogger by revising its LLM prompt to elicit predictions of several log statements, resulting in a new variant that we designate as \textit{Multi-SCLogger} and serves as the baseline for the multi-log setting. To demonstrate the generalizability of our approach, we further conduct experiments on three mainstream LLMs: DeepSeek-V3~\cite{deepseek2024chat}, LLaMA3-70B~\cite{meta2024llama3}, and OpenAI o3-mini~\cite{openai2024gpt4o}. 

\vspace{4pt} \noindent \textbf{Log Position Evaluation Metrics.} For single-log generation setting, consistent with \cite{b17}, we use Position Accuracy (PA) to assess log-position prediction. A prediction is counted as correct (PA = 1) if the predicted line number deviates from the true line number by at most one and both lines are located within the same code block; otherwise, PA = 0. For the multi-log setting, we use precision, recall, and F1 score to evaluate the accuracy and to measure the proportion of false positives and true positives.

\vspace{4pt} \noindent \textbf{Log Level Evaluation Metrics.} We employ L-ACC (level accuracy) and Average Ordered Distance (AOD) to measure log-level prediction, as in~\cite{b6,b7,b15}. L-ACC represents the proportion of correctly predicted levels. AOD captures the ordinal distance between levels, acknowledging that log levels are not independent categories (e.g., \textit{error} is closer to \textit{warn} than to \textit{trace}). 



\vspace{4pt} \noindent \textbf{Log Variables Evaluation Metrics.} We adopt precision, recall, and F1-score to evaluate the set of variables referenced in the generated log statements. For each generated log statement, we denote the set of variables predicted by the model as $S_p$, and the set of variables in the ground-truth log as $S_g$. The evaluation metrics are defined as follows: Precision $= \frac{|S_p \cap S_g|}{|S_p|}$, Recall $= \frac{|S_p \cap S_g|}{|S_g|}$, F1 $= \frac{2 \cdot \text{Precision} \cdot \text{Recall}}{\text{Precision} + \text{Recall}}$. It is important to note that if a predicted variable has the same name as a ground-truth variable but differs in the usage of its member function, it is still considered an incorrect prediction.

\vspace{4pt} \noindent \textbf{Log Messages Evaluation Metrics.} Following prior studies, we employ \textsc{BLEU}--$K$ ($K=\{1,4\}$)~\cite{b32} and \textsc{ROUGE}--$K$ ($K=\{1,L\}$)~\cite{b33} to quantify surface-level similarity, and additionally adopt \textsc{BERTScore} to capture semantic similarity, because \textsc{BLEU} and \textsc{ROUGE} operate solely on $K$-gram overlap. \textsc{BERTScore} ranges from~0 to~1, with higher values indicating better quality.

\vspace{4pt} \noindent \textbf{Experimental Environment.}~\label{sect:implementation} The static analysis component of \codename is implemented with 4,293 lines of Java and shell code, combining Joern~\cite{b42} and Eclipse JDT Core~\cite{b55} to enable comprehensive analysis of Java source code. All experiments for \codename and the baselines are executed on a Linux machine running Ubuntu 22.04.5 LTS, equipped with an AMD EPYC 9354 32-Core Processor @ 3.8 GHz, 2 NVIDIA L40S GPUs (each with 46GB of memory), and 256 GB of RAM. We employ the public API to access O3-mini-20240131 and Deepseek-V3, while LLaMA3-70B-Chat-HF is executed locally. To ensure deterministic outputs and facilitate stability evaluation in log generation, we set the temperature to 0.

\subsection{RQ1: Single-Log Generation Scenario}

\begin{table*}[h]
\caption{Results in the Single-Log Prediction Setting.}
\vspace{-0.1in}
\begin{center}
\resizebox{0.9\textwidth}{!}{
\begin{tabular}{|c|c|c|c|c|c|c|c|c|c|c|c|}
\hline
\multirow{2}{*}{\textbf{Model}}&    \textbf{Position}&\multicolumn{2}{c|}{\textbf{Logging Levels}}&    \multicolumn{3}{c|}{\textbf{Logging Variables}}&\multicolumn{5}{c|}{\textbf{Logging Texts}}\\\cline{2-12}

&     \textbf{PA}&\textbf{L- ACC}&\textbf{AOD}&    \textbf{Precision}&\textbf{Recall}&\textbf{F1}&\textbf{BLEU-1}&\textbf{BLEU-4}&\textbf{ROUGE-1}& \textbf{ROUGE-L} &\textbf{BERTScore}\\
\hline
SCLogger&     0.417&0.573&0.807&    0.535&0.613&0.571&0.474&0.201&0.354&  0.34&0.341\\
\hline
 LANCE&   0.341&0.59&0.747&    0.286&0.302&0.294&0.514&0.184&0.275& 0.273&0.357\\ \hline
 LANCE2.0& 0.383& \textbf{0.613}& 0.775& 0.311& 0.397& 0.349& \textbf{0.544}& 0.182& 0.315& 0.297&0.37\\\hline\hline
 \codename& \textbf{0.54}& 0.609& \textbf{0.815}& \textbf{0.589}& \textbf{0.638}& \textbf{0.607}& 0.514& \textbf{0.235}& \textbf{0.442}& \textbf{0.42}&\textbf{0.467}\\\hline\end{tabular}}
\label{tab1}
\end{center}
\end{table*}

We first evaluate \codename's performance on a dataset constructed by randomly removing only one log statement per method, and compare it with several state-of-the-art baseline techniques. The evaluation results are presented in Table~\ref{tab1}, with the best results for each metric highlighted in bold. 

For log position, \codename outperforms all baselines. Specifically, it improves upon the best-performing baseline, SCLogger~\cite{b28}, by 29.5\% from 0.417 to 0.54. This confirms that our proposed block-type-based prompting strategy is highly effective. In terms of log level prediction, \codename achieves the highest AOD score. Although its accuracy is slightly lower (0.609 vs. 0.613) than that of LANCE 2.0, it outperforms all other baselines. For log variable prediction, \codename surpasses all baselines in accuracy, recall, and F1 score, demonstrating that the function-aware extension indeed improves the accuracy of log variable prediction. Regarding the log message, \codename shows significant improvements over all baseline methods, surpassing the best baseline by 30.80\% (0.467 vs. 0.37) in BERTScore. These results suggest that by extracting semantic dependency information, our system generates log messages that better match the project style and more accurately convey developer intent, making it more suitable for real-world development.

\begin{answerbox}
Answer to RQ1: \codename can significantly outperform state-of-the-art techniques, demonstrating strong effectiveness in the single-log generation setting.
\end{answerbox}

\begin{figure}[h]
    \centering
    \includegraphics[width=\linewidth]{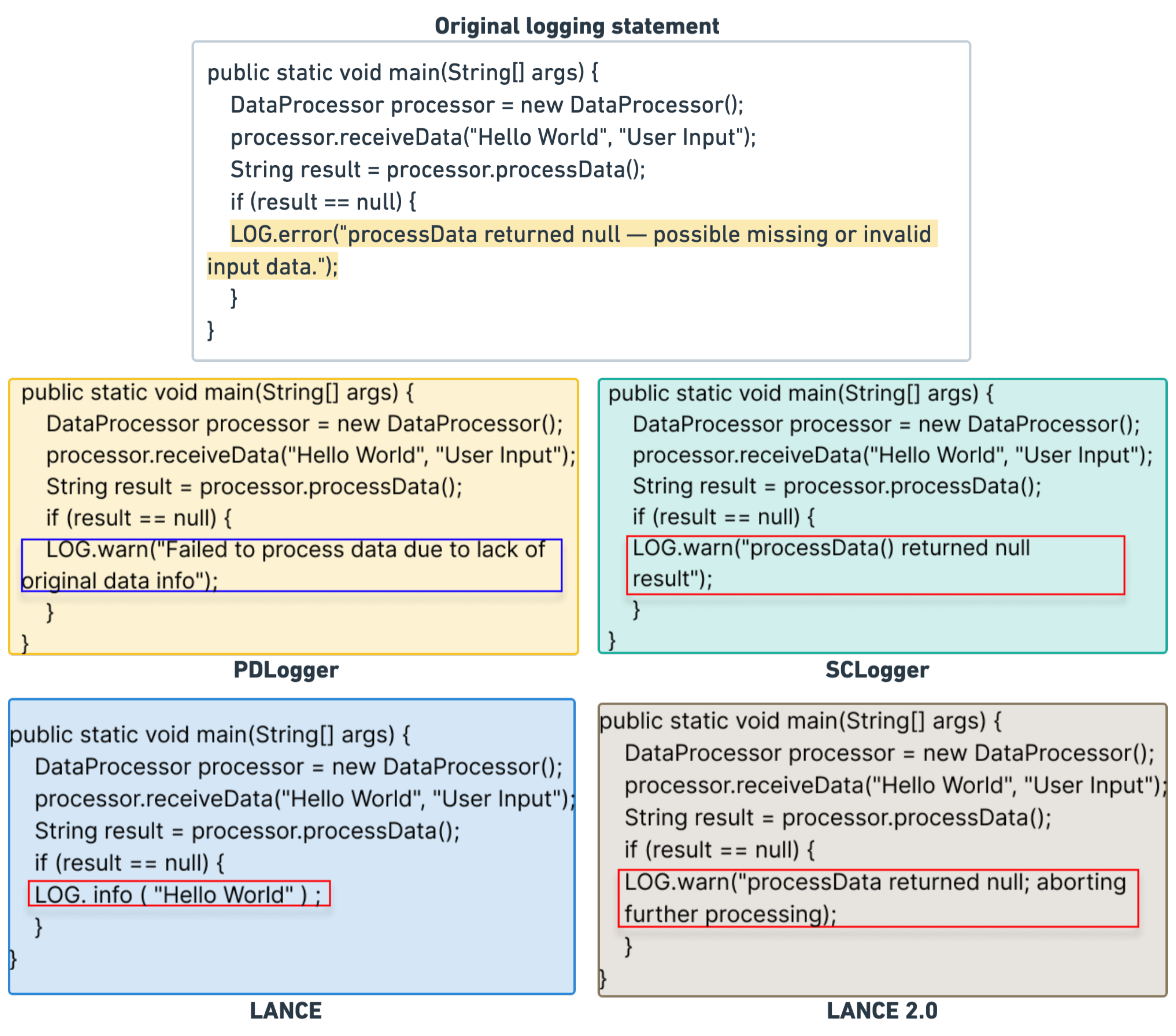}
    \caption{A Case Study in the Single-log Generation Task}
    \vspace{-0.1in}
    \label{fig8}
\end{figure}

\vspace{4pt} \noindent \textbf{Case Study.} Figure~\ref{fig8} presents a case study, with the developer-written log highlighted, to illustrate the effectiveness of \codename in single-log insertion tasks. All four tools correctly identify the appropriate log position. However, only \codename accurately captures the semantic dependencies, enabling it to generate a log message that explains the root cause of \texttt{result == null}. In contrast, SCLogger and LANCE 2.0 merely record the occurrence of the condition, while the message produced by LANCE lacks any diagnostic value. Notably, the original developer-written log also attempts to convey the root cause of \texttt{result == null}, further demonstrating that the log generated by \codename is more consistent with developer intent and can meet practical diagnostic requirements.

\subsection{RQ2: Multi-Log Generation Scenario}

To evaluate \codename's practicality in real-world development, we assess its performance on a dataset where all log statements are removed from methods. The modified SCLogger (\textit{Multi-SCLogger}) serves as the primary baseline. 

\renewcommand{\arraystretch}{1.5}
\begin{table*}[ht]
\caption{Results in the Multiple-Log Prediction Setting.}
\vspace{-0.1in}
\begin{center}
\resizebox{0.9\textwidth}{!}{
\begin{tabular}{|c|c|l|l|c|c|c|c|c|c|c|c|c|c|}
\hline
\multirow{2}{*}{\textbf{Model}}&    \multicolumn{3}{c|}{\textbf{Position}}&\multicolumn{2}{c|}{\textbf{Logging Levels}}&    \multicolumn{3}{c|}{\textbf{Logging Variables}}&\multicolumn{5}{c|}{\textbf{Logging Texts}}\\\cline{2-14}

&     \textbf{Precision}& \textbf{Recall}&\textbf{F1}&\textbf{L- ACC}&\textbf{AOD}&    \textbf{Precision}&\textbf{Recall}&\textbf{F1}&\textbf{BLEU-1}&\textbf{BLEU-4}&\textbf{ROUGE-1}& \textbf{ROUGE-L} &\textbf{BERTScore}\\
\hline
\textit{Multi-SCLogger}&     0.24& \textbf{0.784}&0.367&0.446&0.739&    0.283&0.315&0.28&0.438&0.174&0.384&  0.375&0.324\\\hline\hline
 \codename& \textbf{0.575}& 0.674&\textbf{0.621}& \textbf{0.813}& \textbf{0.935}& \textbf{0.656}& \textbf{0.657}& \textbf{0.657}& \textbf{0.57}& \textbf{0.315}& \textbf{0.487}& \textbf{0.469}&\textbf{0.537}\\\hline\end{tabular}}
\label{tab2}
\end{center}

\end{table*}

As depicted in Table~\ref{tab2}, \codename can vastly outperform \textit{Multi-SCLogger} in all four log component generations. In particular, for log position prediction, we evaluate whether the predicted logs are neither excessive (causing system overhead) nor insufficient (providing too little information). Using precision, recall, and F1 score as metrics, we find that \codename surpasses SCLogger in both precision and F1 score. Specifically, \codename improves F1 score by 69.21\% (0.621 vs. 0.367). This is primarily because \textit{Multi-SCLogger} generates a substantial number of false positives, whereas \codename significantly reduces false positives by employing block-type-based prompting and refinement.

In terms of log level prediction, \codename surpasses SCLogger in both level accuracy (L-ACC) and AOD, with improvements of 82.2\% and 26.5\% respectively. These results indicate that semantic dependency information can also assist \codename in assessing the importance of log statements and predicting appropriate log levels. For log variable prediction, \codename achieves remarkable gains with an F1 score improved by a whopping 135\% (0.657 vs. 0.28). Regarding log message generation, \codename significantly outperforms SCLogger, with a BLEU-4 score of 0.314 (45.4\% higher) and a BERTScore of 0.544 (52.1\% higher). These gains stem from \codename’s effective capture of semantic dependencies, allowing it to better capture developer intent and produce semantically aligned log messages, unlike SCLogger’s heuristics-based random sampling of callers and callees.

\begin{answerbox}
Answer to RQ2: By constructing block-type-based structured prompts and capturing semantic dependencies, \codename significantly outperforms \textit{Multi-SCLogger} across all four dimensions: position, level, variable, and message, in a more practical multi-log generation scanrio.
\end{answerbox}

\begin{figure}[h]
    \centering    \includegraphics[width=\linewidth]{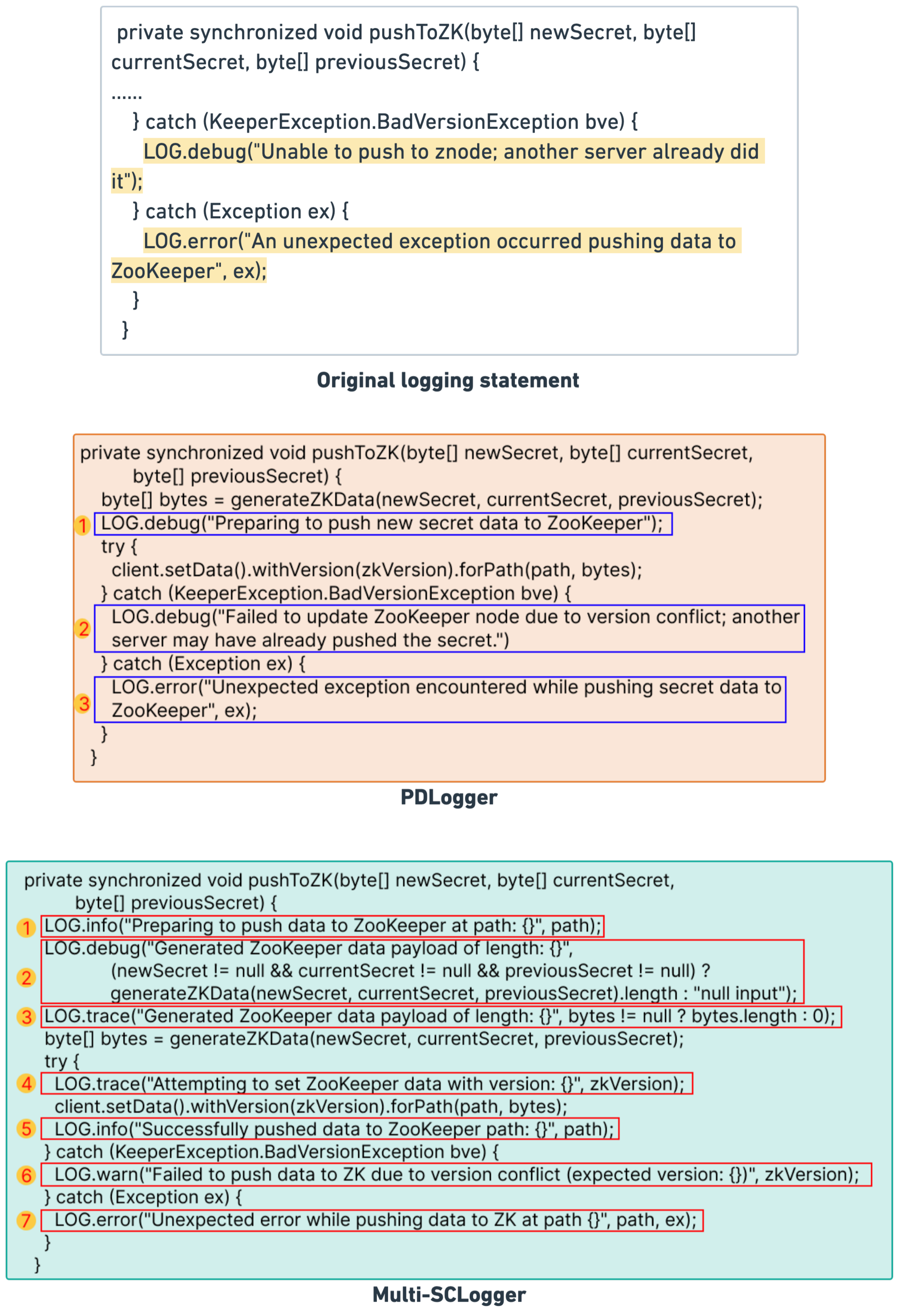}
    \caption{A Case Study in the Multi-Log Generation Scenario}
    \label{fig9}
    \vspace{-0.2in}
\end{figure}

\vspace{4pt} \noindent \textbf{Case Study.} Figure~\ref{fig9} presents a case study illustrating the effectiveness of \codename in the multi-log generation scenario. The ground truths have been highlighted. For log position, both tools achieve 100\% recall. However, \codename’s block-type-based prompting yields only one false positive (66.6\% precision), compared to five from \textit{Multi-SCLogger} (28.6\% precision). Regarding the log message, for the first log, \textit{Multi-SCLogger} generates a superficial message that fails to explain the root cause of the push data failure. In contrast, \codename captures the semantic dependencies and attributes the issue to a prior operation by another server, closely matching the ground truth. Moreover, \codename correctly predicts the log level and variable, while \textit{Multi-SCLogger} does not. For the second log, while both tools generate accurate messages and levels, \textit{Multi-SCLogger} mispredicts one variable ($path$).

\renewcommand{\arraystretch}{1}
\begin{table*}[t]
\setlength{\tabcolsep}{1pt}
\caption{Ablation Study of \codename.}
\vspace{-0.1in}
\begin{center}
\resizebox{\textwidth}{!}{
\begin{tabular}{|c|c|c|c|c|c|c|c|c|c|c|c|c|c|}
\hline

\multirow{2}{*}{\textbf{Model}} & \multicolumn{3}{c|}{\textbf{Position}} 
& \multicolumn{2}{c|}{\textbf{Logging Levels}} 
& \multicolumn{3}{c|}{\textbf{Logging Variables}} 
& \multicolumn{5}{c|}{\textbf{Logging Texts}} \\
\cline{2-14}

&     \textbf{Precision}& \textbf{Recall}&\textbf{F1}&\textbf{L- ACC}&\textbf{AOD}&    \textbf{Precision}&\textbf{Recall}&\textbf{F1}&\textbf{BLEU-1}&\textbf{BLEU-4}&\textbf{ROUGE-1}& \textbf{ROUGE-L} &\textbf{BERTScore}\\
\hline
\codename&     \textbf{0.575}& 0.674&\textbf{0.621}&\textbf{0.813}&\textbf{0.935}&\textbf{0.656}&\textbf{0.657}&\textbf{0.657}&\textbf{0.57}&\textbf{0.315}&0.487&  0.469&0.537\\\hline
\makecell{w/o Block-type-based \\Structured Prompt Construction}  & 0.246& 0.718&0.366& 0.575& 0.798& 0.443& 0.492& 0.466& 0.442& 0.143& 0.332& 0.317&0.473\\\hline
 w/o Semantic-dependency Extension& 0.575& 0.674& 0.621& 0.685& 0.819& 0.413& 0.411& 0.412& 0.482& 0.202& 0.392& 0.374&0.456\\\hline
 w/o Function-aware Extension& 0.575& 0.674& 0.621& 0.606& 0.768& 0.56& 0.559& 0.56& 0.538& 0.253& 0.428& 0.411&0.493\\\hline
 w/o Deduplication& 0.489& \textbf{0.704}& 0.573& 0.813& 0.935& 0.656& 0.657& 0.657& 0.572& 0.314& \textbf{0.498}& \textbf{0.474}&\textbf{0.544}\\\hline
 w/o Log-Level Refinement & 0.575& 0.674& 0.621& 0.733& 0.833& 0.656& 0.657& 0.657& 0.57& 0.315& 0.487& 0.469&0.537\\\hline\end{tabular}}
\label{tab3}
\end{center}

\end{table*}

\subsection{RQ3: Ablation Study}

We further conduct an ablation study to demonstrate the effectiveness of five major design choices in \codename: block-type prompts, semantic-dependency, function-aware extensions, deduplication, and level refinement. We then create five variants, each with one design choice removed\footnote{As semantic slicing depends on block-type prompts, we use the first eligible line in the same block as the slicing entry when block-type prompting is removed.}, and compare them against the full-fledged \codename.

\begin{figure*}[htbp]
  \centering
  \begin{subfigure}[b]{0.4\textwidth}
    \centering
    \includegraphics[width=0.65\textwidth]{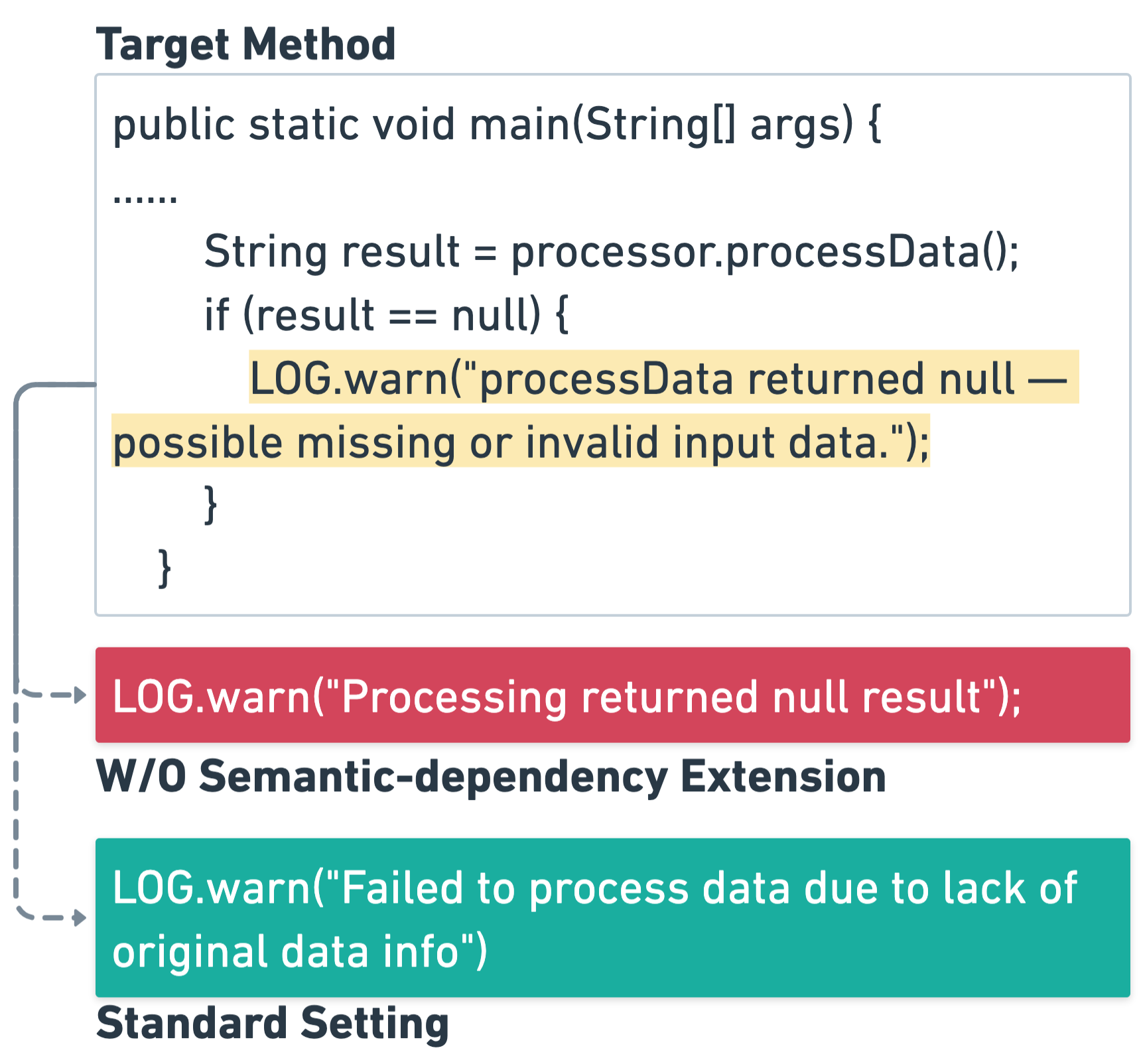}
    \caption{Semantic-dependency Extension Removed}
    \label{ablationstudy:sub1}
  \end{subfigure}
  \begin{subfigure}[b]{0.4\textwidth}
    \centering
    \includegraphics[width=0.8\textwidth]{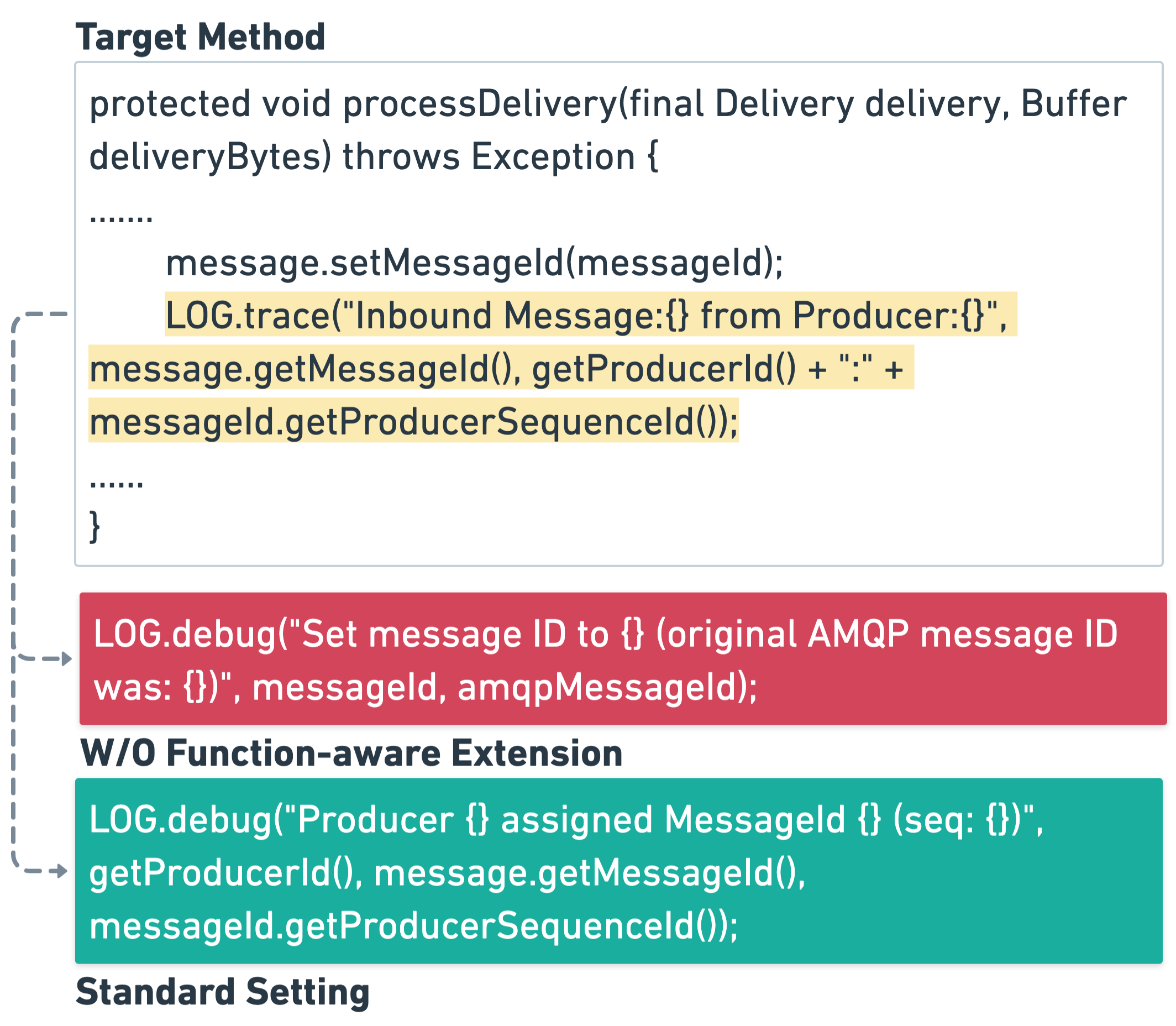}
    \caption{Function-Aware Extension Removed}
    \label{ablationstudy:sub2}
  \end{subfigure}
  \caption{Case Study of the Ablation Study}
  \label{ablationstudy}
\end{figure*}
As shown in Table~\ref{tab3}, removing block-type-based prompting increases the number of false positive positions nearly fourfold. Leading to a sharp decline in precision (-58.3\%) and level accuracy (-29.3\%), primarily due to inaccurate slicing entry points, which result in uninformative or even misleading semantic slices. Removing the semantic dependency extension substantially degrades the quality of log messages. For example, BERTScore drops by 15.1\%, indicating that the extension effectively contributes to generating semantically more aligned log messages. Removing the Function-aware Extension further reduces variable F1 by 14.8\%. Without deduplication, recall for position slightly increases by 4.3\%, but precision drops by 17.59\%, leading to an overall F1 score decrease of 8.38\%, confirming its effectiveness in suppressing false positives. Lastly, excluding level refinement decreases level accuracy by 9.8\%, showing its value in predicting appropriate log levels.

\vspace{4pt} \noindent \textbf{Case Study.} Figure~\ref{ablationstudy} presents two cases that demonstrate how \codename benefits from each phase of the framework. The highlighted lines denote the original log statements. As shown in Figure~\ref{ablationstudy:sub1}, without capturing semantic dependency information, the variant fails to identify the root cause that leads to the conditional statement \verb|if (result == null)| being true, and merely generates a generic log indicating that \verb|result| is null. However, after incorporating semantic dependency expansion, \codename can understand that the underlying reason for \verb|result| being null is due to a lack of original data info, thereby enabling the generation of a more informative log message that facilitates and accelerates system debugging during development.

In the case shown in Figure~\ref{ablationstudy:sub1}, without applying function-aware extension to the target method, \codename cannot include relevant functions in the variable candidate list. With an explicitly provided function list, \codename autonomously selects appropriate functions as variables, thereby producing higher-quality logging statements, as shown in the green box.

\codename can leverage log level refinement to correct inappropriate log levels. For example, adjusting an improper \verb|debug| level in \texttt{LOG.debug("Failed to process data due to lack of original data info")} to \verb|error|, to alert developers to a potential failure rather than simply providing debugging information. 

\begin{answerbox}
Answer to RQ3: Ablation results indicate that each major design choice in \codename plays a critical role in the overall effectiveness.
\end{answerbox}

\vspace{-0.1in}
\subsection{RQ4: Generalizability Study}

To evaluate \codename’s generalizability, we deploy it on three widely used and representative LLMs: OpenAI o3-mini, Deepseek-Chat, and LLaMA-3-70B-chat. Note that OpenAI o3-mini is the default backbone used in \codename.

\renewcommand{\arraystretch}{1}
\begin{table*}[t]

\caption{The Performance of \codename and SCLogger with Different Backbone Models.}
\vspace{-0.1in}
\begin{center}
\resizebox{\textwidth}{!}{
\begin{tabular}{|c|c|c|c|c|c|c|c|c|c|c|c|c|c|c|}
\hline
\multirow{2}{*}{\textbf{Model}} & \multirow{2}{*}{\textbf{Approach}} & \multicolumn{3}{c|}{\textbf{Position}} & \multicolumn{2}{c|}{\textbf{Logging Levels}} & \multicolumn{3}{c|}{\textbf{Logging Variables}} & \multicolumn{5}{c|}{\textbf{Logging Texts}} \\
\cline{3-15}
                                &                   & \textbf{Precision} & \textbf{Recall} & \textbf{F1} & \textbf{L-ACC} & \textbf{AOD} & \textbf{Precision} & \textbf{Recall} & \textbf{F1} & \textbf{BLEU-1} & \textbf{BLEU-4} & \textbf{ROUGE-1} & \textbf{ROUGE-L} & \textbf{BERTScore} \\
\hline

\multirow{3}{*}{O3-mini} & \codename & 0.575 & 0.674 & 0.621 & 0.813 & 0.935 & 0.656 & 0.657 & 0.657 & 0.57 & 0.315 & 0.487 & 0.469 & 0.537 \\\cline{2-15}
                                   & SCLogger & 0.24 & 0.784 & 0.367 & 0.446 & 0.739 & 0.283 & 0.315 & 0.28 & 0.438 & 0.174 & 0.384 & 0.375 & 0.324 \\\cline{2-15}
                                   & $\vartriangle$ & 139\% & -16.3\% & 69.2\% & 82.3\% & 19.6\% & 131.8\% & 108.5\% & 134.6\% & 30.1\% & 81\% & 26.8\% & 25.1\% & 65.7\% \\\hline

 \multirow{3}{*}{Llama-3-70b}&\codename& 0.548& 0.646& 0.593& 0.729& 0.853& 0.605& 0.588& 0.597& 0.482& 0.243& 0.408& 0.394&0.485\\\cline{2-15}
  &SCLogger& 0.142& 0.844& 0.243& 0.424& 0.691& 0.416& 0.501& 0.455& 0.41& 0.125& 0.325& 0.319&0.428\\\cline{2-15}
 & $\vartriangle$ & 285.9\%& -23.5\%& 144.0\%& 71.9\%& 23.4\%& 45.4\%& 17.4\%& 31.2\%& 17.6\%& 94.4\%& 25.5\%& 23.5\%&13.3\%\\\hline
 \multirow{3}{*}{Deepseek-chat}&\codename& 0.523& 0.658& 0.583& 0.749& 0.891& 0.463& 0.437& 0.45& 0.468& 0.21& 0.416& 0.394&0.456\\\cline{2-15}
  &SCLogger& 0.16& 0.982& 0.275& 0.428& 0.68& 0.44& 0.49& 0.464& 0.423& 0.137& 0.359& 0.349&0.355\\ \cline{2-15}
 & $\vartriangle$ & 226.8\%& -32.9\%& 112\%& 75\%& 31\%& 5.2\%& -10.8\%& -3.0\%& 10.6\%& 53.3\%& 15.9\%& 12.9\%&28.5\%\\\hline\end{tabular}}
\label{tab4}
\end{center}

\end{table*}

As shown in Table~\ref{tab4}, \codename consistently outperforms SCLogger across all tested models, demonstrating strong generalization capabilities. On average, \codename improves log position prediction F1 score by 108.4\%, log level accuracy by 76.4\%, log variable F1 score by 54.3\%, and log message BERTScore by 60.3\%. Furthermore, models with stronger comprehension capabilities, such as OpenAI o3 mini, exhibit greater performance when integrated with \codename. 

\begin{answerbox}
Answer to RQ4: \codename maintains high effectiveness in log generation when used with different LLMs, showcasing strong generalization capabilities across backbone models.
\end{answerbox}

\section{Discussion}

\vspace{4pt} \noindent \textbf{Practical Implications.} The adoption of an automated logging solution hinges on deployability. \codename automatically injects an appropriate number of high-quality log statements into projects without any logs— functionality absent from prior approaches. Developers input only their source code, and \codename outputs a predicted‑log augmented project with no extra effort. This substantially reduces developers’ workload and mitigates the common problem of “after-the-fact” log insertion. Thus, \codename integrates into workflows at a low cost and high level of automation, offering strong practical value.

\vspace{4pt} \noindent \textbf{Limitations.} Two main limitations arise. (i) Evaluation uses mostly Java projects, leaving cross‑language generality uncertain; yet our technique is language‑agnostic and could transfer with moderate adaptation. (ii) The pipeline struggles when several logs should be placed in one block. Future work should explore reducing false positives in multi-log scenarios.


\vspace{4pt} \noindent \textbf{Threats to Validity.} Given that \codename relies on large language models (LLMs) during processing, it raises data‑leakage risks for proprietary code. Mitigations include asserting code copyright, running LLMs offline, and evaluating leakage risk before adoption in closed‑source projects.
\section{Related Work}


Logging Practices. Software logs are indispensable for understanding system behaviour and diagnosing failures. According to \cite{b18}, logs appear every 30 lines of code on average and improve fault diagnosis efficiency by 2.2×.

\noindent However, \cite{b13} finds that commits involving log insertions are rare in version-control history, suggesting that most logs are added retrospectively. Recent work \cite{b20} shows that large language models (LLMs) are effective for automatic log-statement generation. A key challenge is balancing log quantity: too many logs cause runtime overhead \cite{b21}, while too few risk missing critical information \cite{b22}, hindering diagnosis. To bridge this gap, this paper proposes the \codename.


Automated Logging Statements. Research on automated logging traditionally divides the task into predicting the log position, level, message and variable. For log-position prediction. A variety of methods seek to identify suitable insertion position \cite{b26,b12,b23,b24}. While \cite{b26} learns developers’ habits from existing repositories, it does not incorporate rich contextual information. \cite{b23} observes that logging decisions depend on both syntactic structure and semantic context, yet its contextual modelling excludes inter-procedural or cross-method information. 

For log-level recommendation, severity prediction is addressed by \cite{b6,b7}, which employ machine-learning techniques to recommend appropriate log levels. For log-message generation, high-quality, context-aware messages are produced by \cite{b10,b34}. For log-variable selection, variable-selection strategies are explored in \cite{b35}.

Although component-wise methods have advanced the field, they lack end-to-end pipelines. Recent holistic systems \cite{b5,b17,b28,b29,b30} generate complete logs but have limitations: \cite{b17} offers a Java method-level solution with limited context, and \cite{b28} extends static context but lacks semantic dependencies and variable handling. Existing end-to-end approaches remain impractical for real-world use. We propose a practical, deployable method that inserts appropriate high-quality logs into initially log-free projects.

\section{CONCLUSION}
In this paper, we introduce \codename, the first automatically log generation approach that is practically applicable to real-world software development. \codename incorporates semantic dependency information and variables within function scope into language models through block-type-based prompt construction. Moreover, it employs deduplication and level refinement strategies to ensure its usability in practical development scenarios. Experimental results demonstrate that \codename outperforms all baseline methods in overall performance and can be effectively adapted to a wide range of LLMs. We believe that \codename can enhance developer productivity and provide valuable insights for researchers in the field of automated log generation.

\nocite{*}
\bibliographystyle{ACM-Reference-Format}
\bibliography{paper}

\end{document}